 \documentclass[pmlr,twocolumn,10pt]{jmlr} 

\usepackage{booktabs}
\usepackage{siunitx}
\usepackage{float}

\newcommand{\equal}[1]{{\hypersetup{linkcolor=black}\thanks{#1}}}

\theorembodyfont{\upshape}
\theoremheaderfont{\scshape}
\theorempostheader{:}
\theoremsep{\newline}

\jmlrvolume{LEAVE UNSET}
\jmlryear{2023}
\jmlrsubmitted{LEAVE UNSET}
\jmlrpublished{LEAVE UNSET}
\jmlrworkshop{Machine Learning for Health (ML4H) 2023} 

\title[EDAC: Efficient Deployment of Audio Classification Models For COVID-19 Detection]{EDAC: Efficient Deployment of Audio Classification Models For COVID-19 Detection}

\author{%
\Name{Andrej Jovanović}\equal{These authors contributed equally, with work done whilst at UoE.} \Email{contact.me.maddox@gmail.com}\\
\Name{Mario Mihaly}\footnotemark[1] \Email{mihaly.mario98@gmail.com}\\
\Name{Lennon Donaldson}\footnotemark[1] \Email{lennon.donaldson@live.co.uk}\\
\addr University of Edinburgh, United Kingdom
}

\begin{document}

\maketitle

\begin{abstract}
The global spread of COVID-19 had severe consequences for public health and the world economy. The quick onset of the pandemic highlighted the potential benefits of cheap and deployable pre-screening methods to monitor the prevalence of the disease in a population. Various researchers made use of machine learning methods in an attempt to detect COVID-19. The solutions leverage various input features, such as CT scans or cough audio signals, with state-of-the-art results arising from deep neural network architectures. However, larger models require more compute; a pertinent consideration when deploying to the edge. To address this, we first recreated two models that use cough audio recordings to detect COVID-19. Through applying network pruning and quantisation, we were able to compress these two architectures without reducing the model's predictive performance.  Specifically, we were able to achieve an $\sim105.76\times$ and an $\sim19.34\times$ reduction in the compressed model file size with corresponding  $\sim1.37\times$ and $\sim1.71\times$ reductions in the inference times of the two models. 
\end{abstract}

\section{Introduction}
\label{sec:intro}
During the COVID-19 pandemic, there was a surge in machine learning (ML) research investigating approaches to detect the disease, with a particular focus on using cough audio recordings as the input data \citep{Brown2020ExploringData, Hamdi2022Attention-basedSound,Agbley2020Wavelet-BasedClassification, Dentamaro2022AUCOBreath}. Researchers proposed that these models would be deployed on the edge (i.e. on mobile devices) to provide a cheaper and quicker prescreening alternative to the typical Polymerase Chain Reaction (PCR) test \citep{Laguarta2020COVID-19Recordings, Chaudhari2020Virufy:Cough, Dentamaro2022AUCOBreath}. Whilst this decentralised paradigm is attractive for its privacy properties, the computational complexity of the model becomes a pertinent consideration, unlike in centralised settings \citep{Liang2021PruningSurvey}. The efficient deployment of ML models on the edge is a well-researched topic, where methods attempt to reduce the size and computational requirements of models without impinging performance; network pruning and quantisation are two such techniques \citep{Liang2021PruningSurvey}. Yet there is little research in their application to COVID-19 detection models, despite their use in other domains and the necessity for practical deployment in this context. 

The main contribution of this paper attempts to address this aforementioned issue.  To do so, we reproduce two uncompressed models from the literature to stand as our baseline for comparison \citep{KingkorMahantaCOVID-19Augmentation,Hamdi2022Attention-basedSound}. We then apply network pruning and quantisation techniques to determine whether we are able to compress the models whilst maintaining the original predictive performance. Through compressing the networks in this way, the computational requirements needed to deploy these models are significantly reduced. As such, we increase both the likelihood of these COVID-19 detection models winning the hardware lottery \citep{Hooker2020TheLottery}, and the feasibility of their deployment. We hope that this work will shed insight on the practicality and pertinence of neural network compression for disease prediction models in a decentralised setting. We release all materials on \href{https://github.com/EDAC-ML4H/EDAC-ML4H}{GitHub}.

\section{Related Work}
With the impetus of the pandemic, a variety of deep learning techniques have been applied to detect the disease from cough recordings. \citet{Bagad2020CoughSounds, Agbley2020Wavelet-BasedClassification, Nessiem2020DetectingNetworks, Laguarta2020COVID-19Recordings, KingkorMahantaCOVID-19Augmentation} used CNNs to detect the disease using spectrograms or scalograms, an image-based representation of the audio cough signal, as input features. \citet{Hassan2020COVID-19Networks,Pahar2021COVID-19Recordings} applied an LSTM to the problem whilst \citet{Hamdi2022Attention-basedSound} merged a CNN and an LSTM with an attention mechanism. However, a common thread in these studies is that deployment practicalities were not considered. Many of these deep learning (DL) models are very computationally intensive making their deployment to edge devices infeasible. \citet{Rao2021COVID-19Algorithms} did address this concern through pruning their detection model. However, their model was less performant than the state-of-the-art models at the time, and the paper did not apply any quantisation techniques. 

Empirical evidence shows that neural network pruning is able to reduce the size of models whilst having little to no effect on the predictive performance \citep{Han2015LearningNetworks, Suzuki2020SpectralError}. The same effects are seen with quantisation \citep{NagelAQuantization, VanhouckeImprovingCPUs}, where quantisation provides the additional benefit of reducing the time taken for a full forward pass through the model.

\section{Datasets and Preprocessing}
For this study, we make use of two publicly available, crowdsourced datasets comprising audio recordings of participant vocal sounds, the \textbf{Coswara} and \textbf{CoughVid} datasets \citep{SharmaCoswara-ADiagnosis, Orlandic2021TheAlgorithms}. The Coswara dataset, after the preprocessing described in \appendixref{sec:data}, contains 1984 COVID-19 negative and 681 COVID-19 positive samples. Samples are divided into training (70\%), validation (15\%) and test (15\%) sets, following standard industry practice. Every positive sample was augmented to create two new positive samples, reducing the class imbalance. We ensured that the test set was augmented after the datasets were split in order to avoid data leakage; this would inflate model performance on the held-out datasets if unaddressed. The augmentation resulted in 1951 negative and 2034 positive samples across the three sets after removing silent recordings. The recordings were zero-padded or trimmed to 7 seconds (154350 samples at a sample rate of 22050Hz), before extracting the first 15 MFCCs with a frame size of 2048 and a hop size of 512. This resulted in inputs with dimension (\(15 \times 302\)).

\citet{Hamdi2022Attention-basedSound} published a subset of the CoughVid dataset that contains 11624 recordings with 8958 negative and 2666 positive samples. Each recording was trimmed or padded to 156027 samples and mel-spectrograms were extracted (see \appendixref{sec:data} for details) resulting in an input size of (\(39 \times 88 \times 3\)). This dataset was then split 60:20:20 between training, validation, and test sets; the split was chosen  to match our test set size with the validation set size used in \citet{Hamdi2022Attention-basedSound}. To address this class in-balance, augmentation techniques were once again implemented to each set resulting in 17916 and 15996 negative and positive samples, respectively, across the three sets.  As with the Coswara data, the test set was augmented after splitting to avoid data leakage.

\section{Methods}

\subsection{Models}\label{sec:models}
In this work, we make use of two different model architectures. The first is the CNN-based model from \cite{KingkorMahantaCOVID-19Augmentation} The exact model specifications can be found in \appendixref{sec:training}. This architecture was trained and evaluated on the Coswara dataset as per the original paper, using the Adam optimizer with an initial learning rate of \( 1 \times 10^{-4}\) for 200 epochs and a batch size of 32. Binary cross entropy was the chosen loss function. The second architecture was an attention based CNN-LSTM model from \citet{Hamdi2022Attention-basedSound} (see \appendixref{sec:training} for more details). This model was trained and evaluated on the CoughVid dataset as per the original paper,  with the Adamax optimizer with an initial learning rate of $1 \times 10^{-3}$ for 100 epochs, with a batch size of 256. Binary cross entropy was used as the loss function.

\subsection{Pruning}\label{sec:pruning}
Pruning involves identifying model parameters that do not contribute to the performance on a given task. Magnitude-based pruning (MBP), a popular pruning method with impressive empirical performance, determines the importance of a parameter from a trained model according to the parameter's norm; the norm and its importance are directly proportional \citep{Zhu2017ToCompression}.
Once identified, the parameters with the smallest norm are set to zero first according to a predefined sparsity proportion with a minimal increase in the model's loss.   This algorithm is applied for a pre-defined number of pruning steps which occur within a fine-tuning (retraining) period \citep{Blalock2020WHATPRUNING}. In this work, we implement a constant and a polynomial decay pruning schedule, where the former sets a constant sparsity across all pruning steps, whilst the latter iteratively increases the sparsity \citep{Zhu2017ToCompression}. 

\subsection{Quantisation}
Quantisation is a popular method that is used to reduce the necessary computation for hosting deep neural networks (DNNs). Despite the fact that most models are trained with 32-bit floating point precision weights \citep{Gholami2021AInference}, it has been shown that this extra precision is superfluous. There is minimal degradation to model accuracy if lower precision is used. Additionally, this procedure reduces both the memory overhead to store the model and the inference time. In our work, we implement post-training quantisation (PTQ), where the weights of a given pre-trained model are reduced to a lower precision in a zero-shot fashion according to the min-max method  \citep{Liang2021PruningSurvey}. Although it has been shown that simulating a quantised aware training environment is better at preserving model performance \citep{Krishnamoorthi2018QuantizingWhitepaper}, PTQ proved to be sufficient in our experiments.

\section{Evaluation}
\begin{figure*}[htb!]
\begin{center}
\centerline{\includegraphics[width=\textwidth, height=90pt]{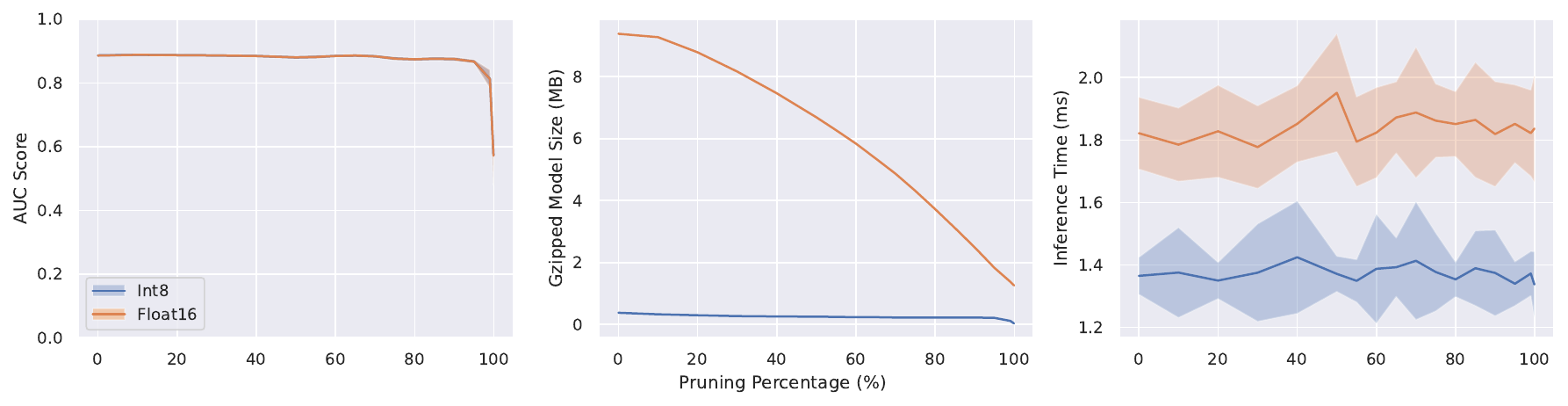}}
\caption{CNN model pruning and quantisation experiment. Figures show the relationship between pruning percentage and model performance, size and inference time, respectively, for two precision values.}
\label{fig:cnn-quant}
\end{center}
\vskip -5mm
\end{figure*}

\begin{figure*}[htb!]
\begin{center}
\centerline{\includegraphics[width=\textwidth, height=90pt]{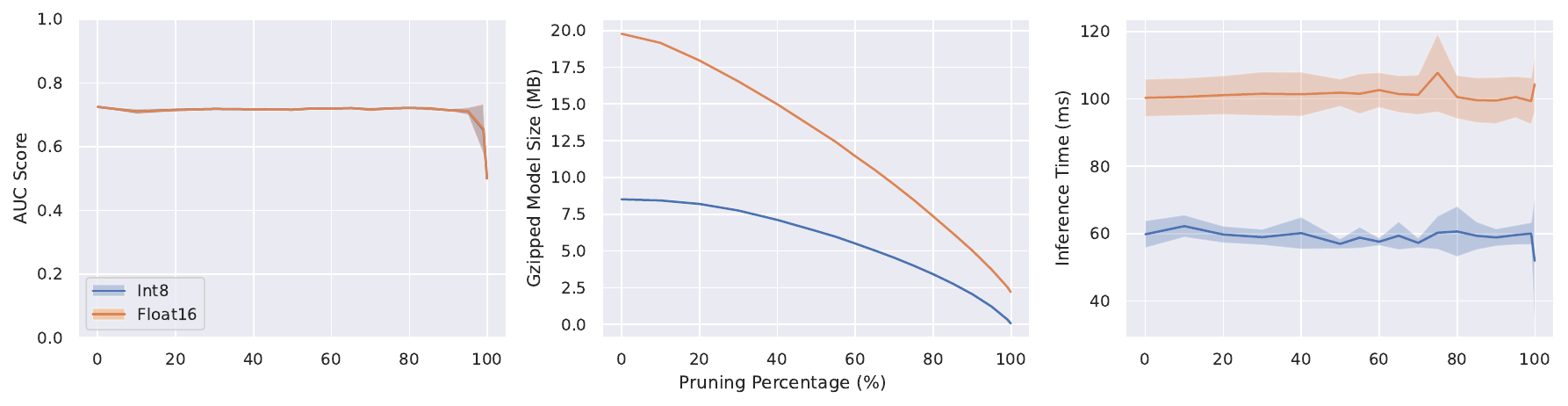}}
\caption{CNN-LSTM attention model pruning and quantisation experiment. Figures show the relationship between pruning percentage and model performance, size and inference time, respectively, for two precision values.}
\label{fig:LSTM-quant}
\end{center}
\vskip -5mm
\end{figure*}
In order to conduct our experiments, we reproduced two performant, uncompressed models from the literature to provide our baseline metrics following the training regime stated in \sectionref{sec:models} (see \appendixref{sec:reproducibility} for more details), where both models were trained on their respective training sets. The baseline results can be seen in \appendixref{sec:reproducibility}. Subsequently, we prune the weights using a 10-epoch fine-tuning process \citep{Blalock2020WHATPRUNING}, as described in \sectionref{sec:pruning}. Performance tracking during this retraining phase relies on the validation set. For all models, including the uncompressed baselines, we calculate the compressed model size, average inference time, and AUC score on the test set. This allows us to analyze the impact of compression on model performance. It is important to note that the primary focus of this paper is not to advance the state-of-the-art or compare the performance of the two models. All experiments were repeated over a number of random seeds to confirm the consistency of our findings (see \appendixref{sec:exps_specs}), where we report the mean and the standard deviation across seeds. All experimental results can be seen in \appendixref{sec:results}.

\subsection{Pruning the Models}\label{sec:pruning_exps}
In our first experiment, we wished to investigate the effect that different levels of magnitude based-pruning (MBP) would have on our baseline model performance. To achieve this, we pruned our baseline models to 16 sparsity proportions (see \appendixref{sec:exps_specs}) using both a polynomial decay and a constant sparsity scheduler. The results of the experiments can be seen in \figureref{fig:cnn-pruning,fig:LSTM-pruning}, and \tableref{tab:cnn_const_sparsity,tab:cnn_poly_decay,tab:cnn_lstm_const_spar,tab:cnn_lstm_poly_decay}.

In the case of both the CNN and CNN-LSTM models, we see that both models can be pruned to 90\% without affecting model performance. After which, there is a significant degradation in model performance. In response to this, the compressed model size decreases approximately linearly in response to the pruning percentage. Compared to the baseline, the CNN model using polynomial decay at 95\% sparsity has an $\sim8.25\times$ reduction in compressed model with a 0.019 point reduction in AUC. The CNN-LSTM has an $\sim5.02\times$ reduction in compressed model size using polynomial decay at 90\% sparsity with a 0.01 point reduction in AUC. However, we find that there is no relationship between inference time and the pruning percentage, with a large amount of variance across random seeds. We posit that this arises as we do not utilise libraries that would leverage the sparsity in our weight matrices when computing a forward pass in the model; the zero values of the pruned weights are considered to be normal weights when vanilla matrix multiplication techniques are applied.

\subsection{Applying Quantisation}\label{sec:quant_exps}
Our second experiment aimed to investigate two things: the trade-off between computational requirements and model performance as a result of quantising our baseline models, and the interaction between quantisation and pruning. However, for brevity, we only visualised the results from polynomial decay and quantisation for both models motivated by polynomial decay's increased stability over constant sparsity. The results can be seen in \figureref{fig:cnn-quant,fig:LSTM-quant}.

For the CNN and CNN-LSTM model, we are able to compress the models significantly without damaging pruning performance as with the pruning experiments (\sectionref{sec:pruning_exps}). Yet, the compression effect is compounded when we use pruning in concert with quantisation with the latter providing enormous benefit in terms of model size and inference time. In the case of the CNN model, we achieve a $\sim105.76\times$ decrease in model size for the 8-bit model relative to the baseline without a large compromise to model performance (0.018 AUC point reduction) with 95\% sparsity. At 90\% sparsity, there was a $\sim19.34\times$ reduction in model size for the 8-bit model relative to the baseline for the CNN-LSTM, whilst losing 0.01 AUC points. In the case of inference time, we see that quantising our models significantly increases the inference speed. At 95\% sparsity, the 8-bit CNN model had a $\sim1.37\times$ reduction in inference time relative to the baseline, whilst the 8-bit CNN-LSTM model had a $\sim1.71\times$ reduction in inference speed relative to its baseline. However, we still observe that there is no relationship between the pruning percentage and inference time due to the lack of sparse matrix multiplications mentioned in \sectionref{sec:pruning_exps}. These experiments show that network compression increases deployability in a decentralised setting.

\section{Discussion}
For successful deployment on the edge, the computational requirements of COVID-19 detection models, and models more generally, have to be taken into consideration. In this work, we show that network compression methods, particularly neural network pruning and quantisation, are able to significantly reduce these requirements, with the best results achieved when applied in concert. This is done without a large comprise to the original performance, increasing the feasibility of the deployment on the edge. 

Further work could extend the findings of this paper through benchmarking our experiments in an on-device setting that supports sparse matrix computation. Additionally, the scope of our paper could be extended to respiratory illness classification more generally. Exploring network compression in this context would be fruitful as the relevant models would be deployed in the same paradigm as the COVID-19 detection models investigated in this paper. Despite the success of the magnitude-based pruning and post-training quantisation, another avenue for exploration could come from exploring the effect of other pruning schemes, such as movement pruning, and quantisation-aware training to the problem. 



\acks{} 
We would like to thank the following individuals for their insightful comments on this manuscript: Bryan M Li, John Skottis, and Simon Chi Lok U.

\bibliography{references}

\begin{thebibliography}{27}
\providecommand{\natexlab}[1]{#1}
\providecommand{\url}[1]{\texttt{#1}}
\expandafter\ifx\csname urlstyle\endcsname\relax
  \providecommand{\doi}[1]{doi: #1}\else
  \providecommand{\doi}{doi: \begingroup \urlstyle{rm}\Url}\fi

\bibitem[Agbley et~al.(2020)Agbley, Li, Haq, Cobbinah, Kulevome, Agbefu, and
  Eleeza]{Agbley2020Wavelet-BasedClassification}
Bless Lord~Y. Agbley, Jianping Li, Aminul Haq, Bernard Cobbinah, Delanyo
  Kulevome, Priscilla~A. Agbefu, and Bright Eleeza.
\newblock {Wavelet-Based Cough Signal Decomposition for Multimodal
  Classification}.
\newblock \emph{2020 17th International Computer Conference on Wavelet Active
  Media Technology and Information Processing, ICCWAMTIP 2020}, pages 5--9, 12
  2020.
\newblock \doi{10.1109/ICCWAMTIP51612.2020.9317337}.

\bibitem[Bagad et~al.(2020)Bagad, Dalmia, Doshi, Nagrani, Bhamare, Mahale,
  Rane, Agarwal, and Panicker]{Bagad2020CoughSounds}
Piyush Bagad, Aman Dalmia, Jigar Doshi, Arsha Nagrani, Parag Bhamare, Amrita
  Mahale, Saurabh Rane, Neeraj Agarwal, and Rahul Panicker.
\newblock {Cough Against COVID: Evidence of COVID-19 Signature in Cough
  Sounds}.
\newblock 9 2020.
\newblock \doi{10.48550/arxiv.2009.08790}.
\newblock URL \url{https://arxiv.org/abs/2009.08790v2}.

\bibitem[Blalock et~al.(2020)Blalock, Ortiz, Frankle, and
  Guttag]{Blalock2020WHATPRUNING}
Davis~W. Blalock, Jose Javier~Gonzalez Ortiz, Jonathan Frankle, and John~V.
  Guttag.
\newblock What is the state of neural network pruning?
\newblock \emph{CoRR}, abs/2003.03033, 2020.
\newblock URL \url{https://arxiv.org/abs/2003.03033}.

\bibitem[Brown et~al.(2020)Brown, Chauhan, Grammenos, Han, Hasthanasombat,
  Spathis, Xia, Cicuta, and Mascolo]{Brown2020ExploringData}
Chloë Brown, Jagmohan Chauhan, Andreas Grammenos, Jing Han, Apinan
  Hasthanasombat, Dimitris Spathis, Tong Xia, Pietro Cicuta, and Cecilia
  Mascolo.
\newblock {Exploring Automatic Diagnosis of COVID-19 from Crowd-sourced
  Respiratory Sound Data}.
\newblock 11, 2020.
\newblock \doi{10.1145/3394486.3412865}.
\newblock URL \url{https://doi.org/10.1145/3394486.3412865}.

\bibitem[Chaudhari et~al.(2020)Chaudhari, Jiang, Fakhry, Han, Xiao, Shen, and
  Khanzada]{Chaudhari2020Virufy:Cough}
Gunvant Chaudhari, Xinyi Jiang, Ahmed Fakhry, Asriel Han, Jaclyn Xiao, Sabrina
  Shen, and Amil Khanzada.
\newblock {Virufy: Global Applicability of Crowdsourced and Clinical Datasets
  for AI Detection of COVID-19 from Cough}.
\newblock 11 2020.
\newblock \doi{10.48550/arxiv.2011.13320}.
\newblock URL \url{https://arxiv.org/abs/2011.13320v4}.

\bibitem[Dentamaro et~al.(2022)Dentamaro, Giglio, Impedovo, Moretti, and
  Pirlo]{Dentamaro2022AUCOBreath}
Vincenzo Dentamaro, Paolo Giglio, Donato Impedovo, Luigi Moretti, and Giuseppe
  Pirlo.
\newblock {AUCO ResNet: an end-to-end network for Covid-19 pre-screening from
  cough and breath}.
\newblock \emph{Pattern Recognition}, 127, 7 2022.
\newblock ISSN 00313203.
\newblock \doi{10.1016/J.PATCOG.2022.108656}.

\bibitem[Gholami et~al.(2021)Gholami, Kim, Dong, Yao, Mahoney, and
  Keutzer]{Gholami2021AInference}
Amir Gholami, Sehoon Kim, Zhen Dong, Zhewei Yao, Michael~W. Mahoney, and Kurt
  Keutzer.
\newblock {A Survey of Quantization Methods for Efficient Neural Network
  Inference}.
\newblock 3 2021.
\newblock URL \url{http://arxiv.org/abs/2103.13630}.

\bibitem[Hamdi et~al.(2022)Hamdi, Oussalah, Moussaoui, Saidi, and
  Saidi~mohamedsaidi]{Hamdi2022Attention-basedSound}
Skander Hamdi, Mourad Oussalah, Abdelouahab Moussaoui, Mohamed Saidi, and
  Mohamed Saidi~mohamedsaidi.
\newblock {Attention-based hybrid CNN-LSTM and spectral data augmentation for
  COVID-19 diagnosis from cough sound}.
\newblock \emph{Journal of Intelligent Information Systems}, 59:\penalty0
  367--389, 2022.
\newblock \doi{10.1007/s10844-022-00707-7}.
\newblock URL \url{https://doi.org/10.1007/s10844-022-00707-7}.

\bibitem[Han et~al.(2015)Han, Pool, Tran, and Dally]{Han2015LearningNetworks}
Song Han, Jeff Pool, John Tran, and William~J. Dally.
\newblock {Learning both Weights and Connections for Efficient Neural
  Networks}.
\newblock \emph{Advances in Neural Information Processing Systems},
  2015-January:\penalty0 1135--1143, 6 2015.
\newblock ISSN 10495258.
\newblock \doi{10.48550/arxiv.1506.02626}.
\newblock URL \url{https://arxiv.org/abs/1506.02626v3}.

\bibitem[Hassan et~al.(2020)Hassan, Shahin, and
  Alsabek]{Hassan2020COVID-19Networks}
Abdelfatah Hassan, Ismail Shahin, and Mohamed~Bader Alsabek.
\newblock {COVID-19 Detection System using Recurrent Neural Networks}.
\newblock \emph{Proceedings of the 2020 IEEE International Conference on
  Communications, Computing, Cybersecurity, and Informatics, CCCI 2020}, 11
  2020.
\newblock \doi{10.1109/CCCI49893.2020.9256562}.

\bibitem[Hooker(2020)]{Hooker2020TheLottery}
Sara Hooker.
\newblock {The Hardware Lottery}.
\newblock \emph{Communications of the ACM}, 64\penalty0 (12):\penalty0 58--65,
  9 2020.
\newblock ISSN 15577317.
\newblock \doi{10.48550/arxiv.2009.06489}.
\newblock URL \url{https://arxiv.org/abs/2009.06489v2}.

\bibitem[Krishnamoorthi(2018)]{Krishnamoorthi2018QuantizingWhitepaper}
Raghuraman Krishnamoorthi.
\newblock {Quantizing deep convolutional networks for efficient inference: A
  whitepaper}.
\newblock 2018.

\bibitem[Laguarta et~al.(2020)Laguarta, Hueto, and
  Subirana]{Laguarta2020COVID-19Recordings}
Jordi Laguarta, Ferran Hueto, and Brian Subirana.
\newblock {COVID-19 Artificial Intelligence Diagnosis Using only Cough
  Recordings}.
\newblock \emph{IEEE Open Journal of Engineering in Medicine and Biology},
  1:\penalty0 275--281, 2020.
\newblock ISSN 26441276.
\newblock \doi{10.1109/OJEMB.2020.3026928}.

\bibitem[Liang et~al.(2021)Liang, Glossner, Wang, Shi, and
  Zhang]{Liang2021PruningSurvey}
Tailin Liang, John Glossner, Lei Wang, Shaobo Shi, and Xiaotong Zhang.
\newblock {Pruning and Quantization for Deep Neural Network Acceleration: A
  Survey}.
\newblock \emph{Neurocomputing}, 461:\penalty0 370--403, 1 2021.
\newblock URL \url{http://arxiv.org/abs/2101.09671}.

\bibitem[Mahanta et~al.(2022)Mahanta, Kaushik, Jain, Truong, and
  Guha]{KingkorMahantaCOVID-19Augmentation}
Saranga~Kingkor Mahanta, Darsh Kaushik, Shubham Jain, Hoang~Van Truong, and
  Koushik Guha.
\newblock Covid-19 diagnosis from cough acoustics using convnets and data
  augmentation, 2022.

\bibitem[Muguli et~al.(2021)Muguli, Pinto, R., Sharma, Krishnan, Ghosh, Kumar,
  Bhat, Chetupalli, Ganapathy, Ramoji, and Nanda]{DiCova2021}
Ananya Muguli, Lancelot Pinto, Nirmala R., Neeraj Sharma, Prashant Krishnan,
  Prasanta~Kumar Ghosh, Rohit Kumar, Shrirama Bhat, Srikanth~Raj Chetupalli,
  Sriram Ganapathy, Shreyas Ramoji, and Viral Nanda.
\newblock Dicova challenge: Dataset, task, and baseline system for covid-19
  diagnosis using acoustics, 2021.

\bibitem[Nagel et~al.(2021)Nagel, Fournarakis, Amjad, Bondarenko, van Baalen,
  and Blankevoort]{NagelAQuantization}
Markus Nagel, Marios Fournarakis, Rana~Ali Amjad, Yelysei Bondarenko, Mart van
  Baalen, and Tijmen Blankevoort.
\newblock A white paper on neural network quantization.
\newblock \emph{CoRR}, abs/2106.08295, 2021.
\newblock URL \url{https://arxiv.org/abs/2106.08295}.

\bibitem[Nessiem et~al.(2020)Nessiem, Mohamed, Coppock, Gaskell, and
  Schuller]{Nessiem2020DetectingNetworks}
Mina~A. Nessiem, Mostafa~M. Mohamed, Harry Coppock, Alexander Gaskell, and
  Bjorn~W. Schuller.
\newblock {Detecting COVID-19 from Breathing and Coughing Sounds using Deep
  Neural Networks}.
\newblock \emph{Proceedings - IEEE Symposium on Computer-Based Medical
  Systems}, 2021-June:\penalty0 183--188, 12 2020.
\newblock ISSN 10637125.
\newblock \doi{10.48550/arxiv.2012.14553}.
\newblock URL \url{https://arxiv.org/abs/2012.14553v1}.

\bibitem[Orlandic et~al.(2021)Orlandic, Teijeiro, and
  Atienza]{Orlandic2021TheAlgorithms}
Lara Orlandic, Tomas Teijeiro, and David Atienza.
\newblock {The COUGHVID crowdsourcing dataset, a corpus for the study of
  large-scale cough analysis algorithms}.
\newblock \emph{Scientific Data}, 8\penalty0 (1), 12 2021.
\newblock ISSN 20524463.
\newblock \doi{10.1038/s41597-021-00937-4}.

\bibitem[Pahar et~al.(2021)Pahar, Klopper, Warren, and
  Niesler]{Pahar2021COVID-19Recordings}
Madhurananda Pahar, Marisa Klopper, Robin Warren, and Thomas Niesler.
\newblock {COVID-19 cough classification using machine learning and global
  smartphone recordings}.
\newblock \emph{Computers in Biology and Medicine}, 135:\penalty0 104572, 8
  2021.
\newblock ISSN 0010-4825.
\newblock \doi{10.1016/J.COMPBIOMED.2021.104572}.

\bibitem[Park et~al.(2019)Park, Chan, Zhang, Chiu, Zoph, Cubuk, and
  Le]{SpecAugment}
Daniel~S. Park, William Chan, Yu~Zhang, Chung-Cheng Chiu, Barret Zoph, Ekin~D.
  Cubuk, and Quoc~V. Le.
\newblock {SpecAugment}: A simple data augmentation method for automatic speech
  recognition.
\newblock In \emph{Interspeech 2019}. {ISCA}, sep 2019.
\newblock \doi{10.21437/interspeech.2019-2680}.
\newblock URL \url{https://doi.org/10.21437%2Finterspeech.2019-2680}.

\bibitem[Rao et~al.(2021)Rao, Narayanaswamy, Esposito, Thiagarajan, and
  Spanias]{Rao2021COVID-19Algorithms}
Sunil Rao, Vivek Narayanaswamy, Michael Esposito, Jayaraman~J Thiagarajan, and
  Andreas Spanias.
\newblock {COVID-19 detection using cough sound analysis and deep learning
  algorithms}.
\newblock \emph{Intelligent Decision Technologies}, 15:\penalty0 655--665,
  2021.
\newblock \doi{10.3233/IDT-210206}.

\bibitem[Sharma et~al.(2020)Sharma, Krishnan, Kumar, Ramoji, Chetupalli, R.,
  Ghosh, and Ganapathy]{SharmaCoswara-ADiagnosis}
Neeraj Sharma, Prashant Krishnan, Rohit Kumar, Shreyas Ramoji, Srikanth~Raj
  Chetupalli, Nirmala R., Prasanta~Kumar Ghosh, and Sriram Ganapathy.
\newblock Coswara {\textemdash} a database of breathing, cough, and voice
  sounds for {COVID}-19 diagnosis.
\newblock In \emph{Interspeech 2020}. {ISCA}, oct 2020.
\newblock \doi{10.21437/interspeech.2020-2768}.
\newblock URL \url{https://doi.org/10.21437%2Finterspeech.2020-2768}.

\bibitem[Suzuki et~al.(2020)Suzuki, Abe, Murata, Horiuchi, Ito, Wachi, Hirai,
  Yukishima, and Nishimura]{Suzuki2020SpectralError}
Taiji Suzuki, Hiroshi Abe, Tomoya Murata, Shingo Horiuchi, Kotaro Ito, Tokuma
  Wachi, So~Hirai, Masatoshi Yukishima, and Tomoaki Nishimura.
\newblock {Spectral Pruning: Compressing Deep Neural Networks via Spectral
  Analysis and its Generalization Error}.
\newblock 2020.

\bibitem[Vanhoucke et~al.(2011)Vanhoucke, Senior, and
  Mao]{VanhouckeImprovingCPUs}
Vincent Vanhoucke, Andrew Senior, and Mark~Z. Mao.
\newblock Improving the speed of neural networks on cpus.
\newblock In \emph{Deep Learning and Unsupervised Feature Learning Workshop,
  NIPS 2011}, 2011.

\bibitem[Xie et~al.(2021)Xie, Zhao, Qiang, Mi, Tang, and Li]{Xie2021}
Yuan Xie, Jisheng Zhao, Baohua Qiang, Luzhong Mi, Chenghua Tang, and Longge Li.
\newblock {Attention Mechanism-Based CNN-LSTM Model for Wind Turbine Fault
  Prediction Using SSN Ontology Annotation}.
\newblock \emph{Wireless Communications and Mobile Computing}, 2021:\penalty0
  1--12, 3 2021.
\newblock ISSN 1530-8677.
\newblock \doi{10.1155/2021/6627588}.

\bibitem[Zhu and Gupta(2017)]{Zhu2017ToCompression}
Michael Zhu and Suyog Gupta.
\newblock {To prune, or not to prune: exploring the efficacy of pruning for
  model compression}.
\newblock 10 2017.
\newblock URL \url{http://arxiv.org/abs/1710.01878}.

\end{thebibliography}

\appendix
\clearpage
\section{Data Preprocessing}\label{sec:data}
\subsection{Coswara}

\citet{KingkorMahantaCOVID-19Augmentation} developed the CNN-based neural network for the DiCOVA 2021 challenge \citep{DiCova2021}, where the dataset was derived from the Coswara dataset. The original dataset used for the challenge is not publicly available, therefore, the rest of this section describes the steps taken to create an approximation of the original dataset.

The full Coswara dataset contains recordings from 2745 participants and includes examples of breathing, coughing, sustained vowel sounds and counting. From the available sounds, the \texttt{shallow} cough recordings were selected. The metadata contains medical information declared by the participants, including the \texttt{Covid Status} field, selected as the label for the classification task. The \textit{\texttt{healthy, no respiratory illness exposed, respiratory illness not identified and recovered fully}} labels were merged under \texttt{Non-Likely-COVID-19 (negative)} while the \textit{\texttt{positive mild, positive moderate and positive asymptomatic}} labels were merged under \texttt{Likely-COVID-19 (positive)}. This yielded a dataset with 1984 negative and 681 positive instances.

Given the class imbalance present in the DiCOVA dataset \citep{DiCova2021}, \citet{KingkorMahantaCOVID-19Augmentation} used the Python Audiomentations\footnote{\href{https://pypi.org/project/audiomentations/}{https://pypi.org/project/audiomentations/}} package to perform data augmentation to improve the class imbalance from 1:12 to 1:3. Similarly, in the reproduced dataset, a class imbalance of 1:3 was present, which was mitigated by augmenting each of the positive instances with two different augmentation configurations utilising the \texttt{TimeStretch, PitchShift, Shift, Trim and Gain} functions from the Audiomentations package. The exact augmentation settings are shared in the attached GitHub repository. The augmented dataset consists of 1951 negative and 2034 positive instances after removing silent recordings.

\subsection{CoughVid}


The attention based CNN-LSTM model was trained on the publicly available CoughVid dataset of over 25000 crowdsourced cough recordings \citep{Orlandic2021TheAlgorithms}. The metadata of the dataset contains self-reported geographical and medical information, including the Covid-19 \texttt{status} field. The metadata further includes a \texttt{cough detected} field, which is the probability of a cough being present in the recording. The latter feature was derived from an XGBoost classifier developed for the task.

\citet{Hamdi2022Attention-basedSound} published a subset of the CoughVid dataset that contains only samples with the Covid-19 \texttt{status} field provided. The recordings were preprocessed using the Unsilence\footnote{\href{https://pypi.org/project/unsilence/}{https://pypi.org/project/unsilence/}} Python library to remove all silence within the recordings, retaining only the most important vocal patterns. The authors identified $P_\epsilon=0.7$ as the optimal probability for \texttt{cough detected}, yielding 11624 recordings with 8958 \texttt{healthy}, 1935 \texttt{symptomatic} and 731 \texttt{COVID} positive samples.

Given the class imbalance, the authors merged the \texttt{symptomatic} and \texttt{COVID} labels under \texttt{Likely-COVID-19} and left the \texttt{Non-Likely-COVID-19 (healthy)} unchanged, turning the problem into a binary classification task.

To further improve the class imbalance, a two-step augmentation pipeline was adopted. In the first step, each of the \texttt{Likely-COVID-19} samples were \texttt{Pitch Shifted} down by four steps ($n\_step=-4$) using the Librosa\footnote{\href{https://librosa.org/doc/latest/index.html}{https://librosa.org/doc/latest/index.html}} Python package. The second step involves the \texttt{SpecAugment} \citep{SpecAugment} three-step augmentation method, which uses mel-spectrograms with frequency and time masking. At this stage, the recordings were standardised to a length of 156027 (7.07 seconds at a sample rate of 22050Hz) using zero padding in the beginning and at the end, when needed.

For each recording, mel-spectrograms were extracted and converted to decibel units using the Librosa Python library with the parameters detailed in Table \ref{tab:mel_spec_params}. Using frequency $F=30$ and time $T=30$ masking parameters,  each instance in the \texttt{Likely-COVID-19} class was augmented twice, whilst each instance in the \texttt{Non-Likely-COVID-19} class was augmented once. These augmentations generated new, random mel-spectograms for each class. \tableref{tab:coughvid_labels} details the class imbalances before and after the CoughVid set was augmented. 

\begin{table}[htb!]
    \centering
    \begin{tabular}{l|c}
        \hline
        Parameter & Value \\ \hline
        n\_mel & 128 \\
        hop\_length & 128 \\
        fmax & 8000 \\
        n\_fft & 512 \\
        center & True \\ \hline
        ref & max \\
        top\_db & 80 \\\hline
    \end{tabular}
    \caption{Mel-spectrogram calculation parameters}
    \label{tab:mel_spec_params}
\end{table}

\begin{table*}[h!]
    \centering
    \begin{tabular}{c|c|c|c} 
     & Non-Likely-COVID-19  & Likely-COVID-19 & Total           \\ \hline
       Original sample count &  8958 (77.06\%)	     &  2666 (22.94\%) & 11624 (100\%)   \\ 
       After pitch shifting  &  8958 (62.69\%)      &  5332 (37.31\%) & 14290 (100\%)   \\ 
       After SpecAugment     & 17916 (52.83\%)      & 15996 (47.17\%) & 33912 (100\%)   \\ 
    \end{tabular}
    \caption{Change in class imbalance through the augmentation pipeline for CoughVid}
    \label{tab:coughvid_labels}
\end{table*}

The resultant mel-spectrograms were saved as PNG files, before processing them using the OpenCV\footnote{\href{https://pypi.org/project/opencv-python/}{https://pypi.org/project/opencv-python/}} Python library. The saved mel-spectrogram images were resized to ($88\times39$) and normalized to meet the input requirements of the model ($39\times88\times3$).

\subsection{Augmentation}

The training and validation sets were augmented together following the practice employed by \cite{KingkorMahantaCOVID-19Augmentation} and \cite{Hamdi2022Attention-basedSound}. This allowed the training of baseline models with performance on the validation sets comparable to the one reported in the original papers. Data augmentation before the separation of the training and validation sets leads to inflated metrics. Therefore, to facilitate a fair comparison of the experiment outcomes, a held-out dataset was retained in both cases that were extracted before any augmentation took place.

\section{Model Specifications}\label{sec:training}
\begin{figure*}[h]
\includegraphics[width=\textwidth,height=4cm]{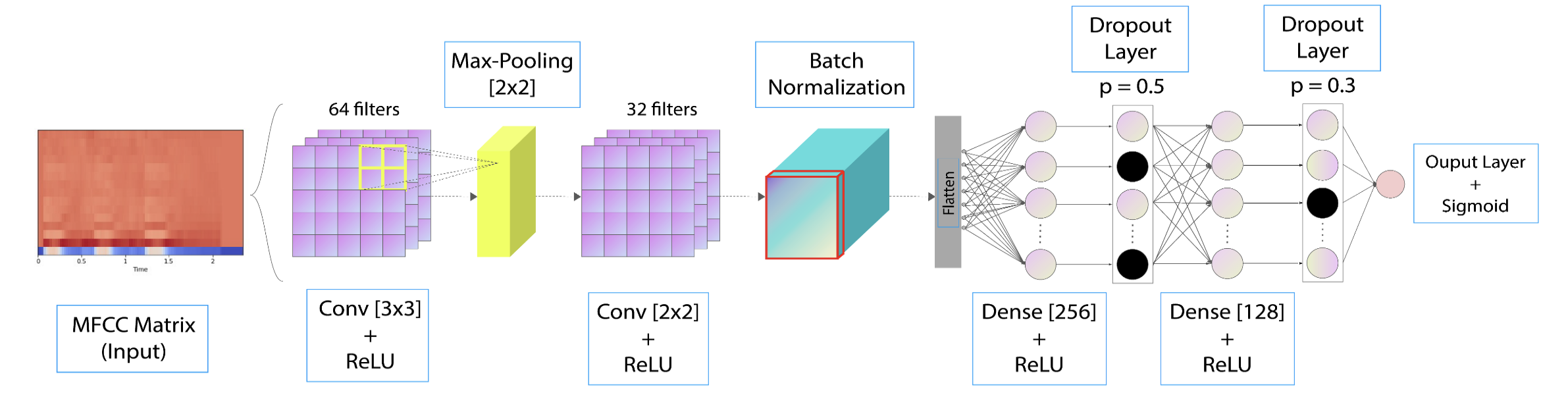}
\caption{Model architecture of baseline model specified in \cite{KingkorMahantaCOVID-19Augmentation}}
\label{fig:brogrammers_fig}
\end{figure*}

The architecture in \cite{KingkorMahantaCOVID-19Augmentation} is outlined in Figure \ref{fig:brogrammers_fig}. In addition to the architecture specifications that are visible in the figure, there are a few notable details:

\begin{enumerate}
    \itemsep0em 
    \item The first fully connected layer has elastic net regularization applied to the layer's weights, with $\lambda_{L1} = 3 \times 10^{-4}, \lambda_{L2} = 4 \times 10^{-3}$ and  L2 regularization is added to the layer's bias with $\lambda = 3\times 10^{-3}$.
    \item Similarly, elastic net regularization is applied to the second fully connected layer's weights with $\lambda_{L1} = 1 \times 10^{-3}, \lambda_{L2} = 1 \times 10^{-2}$ and L2 regularization is added to the bias with $\lambda = 1\times 10^{-2}$.
\end{enumerate}

\begin{figure*}[hbt!]
\includegraphics[width=\textwidth,height=4cm]{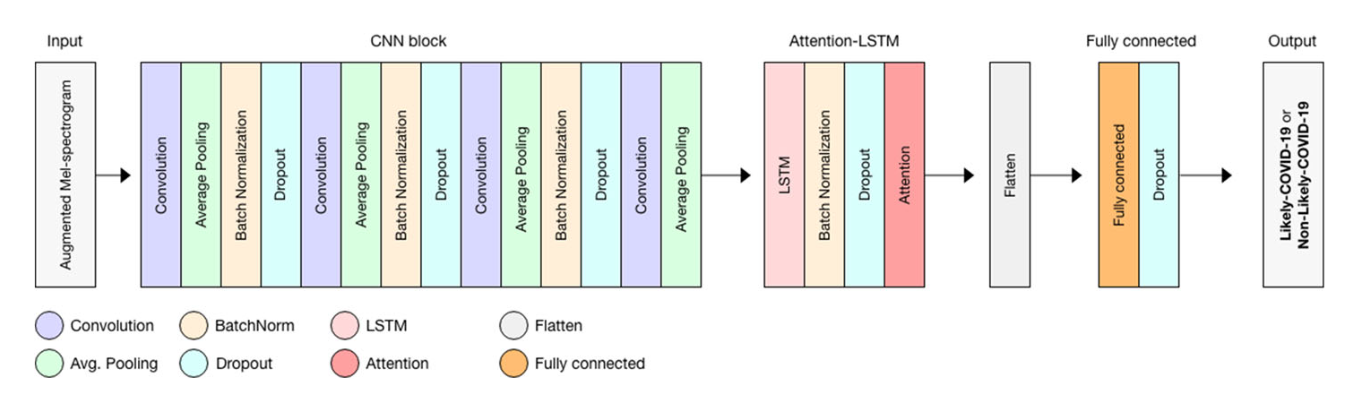}
\caption{Model architecture of baseline model specified in \cite{Hamdi2022Attention-basedSound}}
\label{fig:CNN-LSTM-model}
\end{figure*}

The model takes the $15\times302$ MFCC audio features and outputs a predicted probability $p$ of COVID-19 infection. 
\newline
The second architecture we recreated was the attention based CNN-LSTM model from \cite{Hamdi2022Attention-basedSound}, described in Figure \ref{fig:CNN-LSTM-model}. This was chosen to evaluate the effects of compression on a larger, more complex model \footnote{\href{https://github.com/skanderhamdi}{https://github.com/skanderhamdi}}. When training the model, a dropout probability of 0.2 was used for all layers, apart from the fully-connected layer where a dropout probability of 0.5 was chosen. The model uses the attention mechanism introduced in \cite{Xie2021}. The attention mechanism follows the LSTM layer and focuses on the significant features of the input signal by using a weighted sum of all hidden states, where the weights are derived from a trainable matrix. This results in a contextualised feature vector summarising sequence history. The model receives a mel-spectrogram of shape ($39\times88\times3$) and outputs the predicted probability of the cough indicating a COVID-19-positive user. 

\section{Reproducibility of Initial Baselines}\label{sec:reproducibility}
The reproduced CNN model from \citet{KingkorMahantaCOVID-19Augmentation} achieved an AUC score of 0.886 and the CNN-LSTM model from \citet{Hamdi2022Attention-basedSound} achieved an AUC score of 0.725 on their respective test sets. The reproduced CNN model marginally outperformed the original 0.871 AUC score reported on the DiCova blind test set. This is likely due to using the whole Coswara dataset, giving larger and thus more diverse training and validation sets, leading to the better generalisation of the model.

The AUC score on the test set dropped from 0.911 to 0.725 in the case of the CNN-LSTM model \citet{Hamdi2022Attention-basedSound} employed augmentation before splitting their data into training, validation and test sets. This resulted in information leakage in the test set, resulting in the inflated AUC score reported in the original paper. Despite this, our reproduced results are sufficient to stand as baseline metrics. Additionally, we trained the model for 100 epochs compared to the original 500 epochs as we found that there was no benefit in additional training.

\section{Experiment Specifications}\label{sec:exps_specs}
We repeated the pruning experiments for the CNN model across 20 random seeds and for the CNN-LSTM model across 5 random seeds. The random seeds were chosen to be $1234 + run$, where $run$ corresponds to the enumeration of the reruns starting from 0. The decision to reduce the number of reruns from 20 to 5 for the CNN-LSTM model was motivated by the excessive computational resource requirements of the runs. While the 20 runs for the CNN model combined took just under 8 hours, each run of the CNN-LSTM model took over 14 hours on two T4 GPUs. The inference time was computed by measuring the average time it took for a model to individually classify each sample of the test set. Batch inference was not used as we were interested in the inference time for a single input. Whilst this is not the most robust method to measure inference time (as it relies on background activity of the measuring device remaining constant), it served as a simple metric to investigate potential relationships. The model size was derived from the compressed \textit{tflite} model file size. This was a useful metric as a decrease in the compressed model size would help an edge device in efficiently storing a model's weights. Furthermore, using the compressed model size allowed us to show the effect of TensorFlow’s pruning implementation\footnote{\url{https://www.tensorflow.org/model_optimization}}. Additionally, in the case of the CNN-LSTM model, we did not prune the attention layer as recommended by the TensorFlow model optimisation guide\footnote{\url{https://www.tensorflow.org/model\_optimization/guide/pruning/comprehensive\_guide.md}}.

In our experiment setup, we used the following pruning percentages: 10\%, 20\%, 30\%, 40\%, 50\%, 55\%, 60\%, 65\%, 70\%, 75\%, 80\%, 85\%, 90\%, 95\%, 99\%, 99.9\%. 
\section{Experiment Figures and Tables}\label{sec:results}
\onecolumn
\begin{figure*}[hbt]
\begin{center}
\centerline{\includegraphics[width=\textwidth, height=95pt]{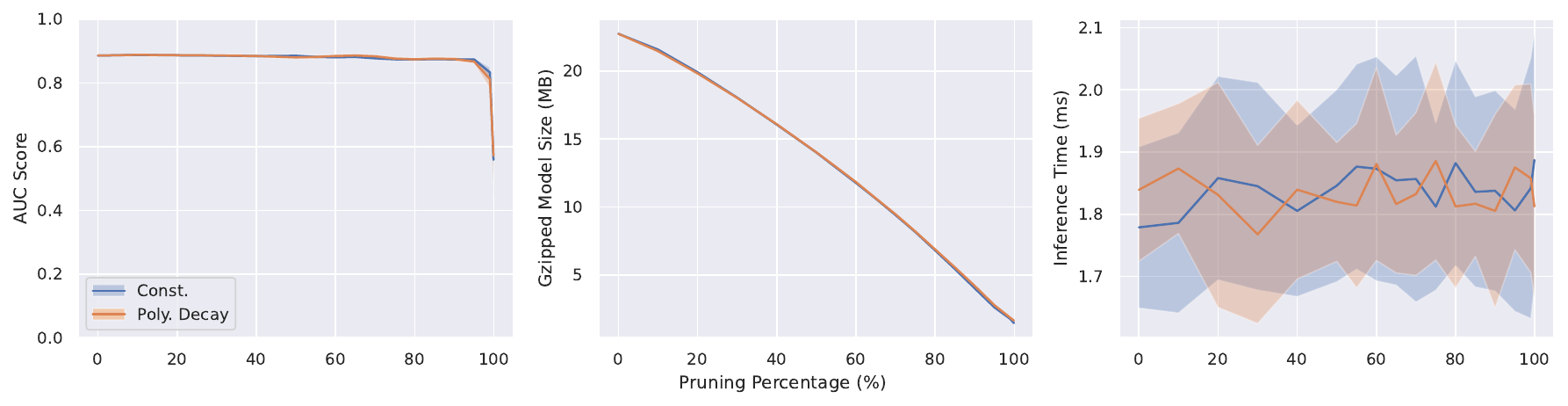}}
\caption{CNN model pruning experiment. Figures show relationship between pruning percentage and model performance, size and inference time, respectively.}
\label{fig:cnn-pruning}
\end{center}
\end{figure*} 
\begin{figure*}[hbt]
\begin{center}
\centerline{\includegraphics[width=\textwidth, height=95pt]{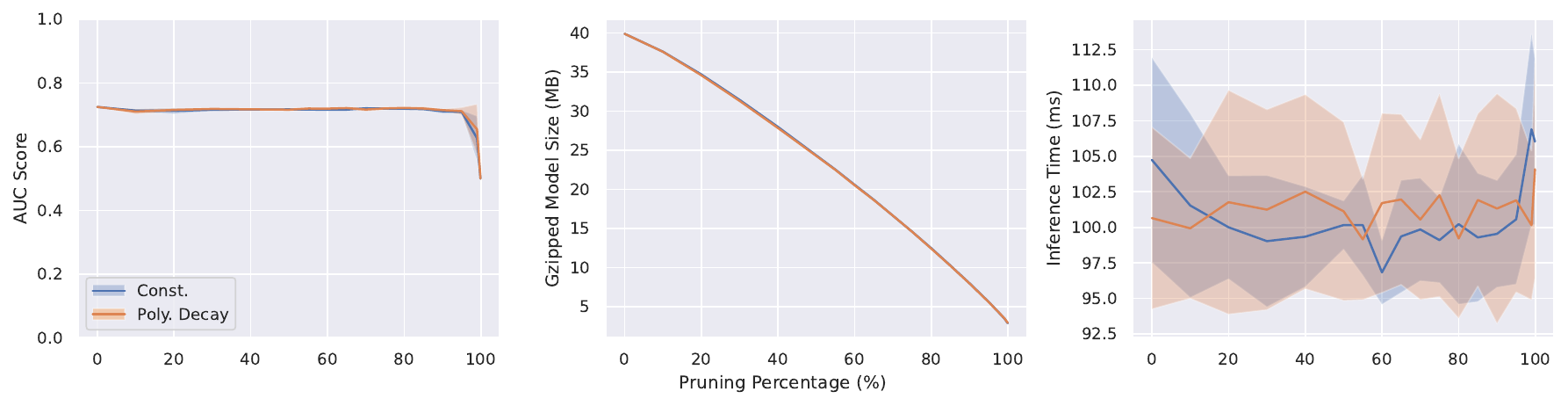}}
\caption{CNN-LSTM attention model pruning experiment. Figures show relationship between pruning percentage and model performance, size and inference time, respectively.}
\label{fig:LSTM-pruning}
\end{center}
\end{figure*}

\begin{table*}[h]
\begin{center}
\begin{small}
\begin{sc}
\scalebox{0.45}{%
\begin{tabular}{cccccccccc}
\hline
Sparsity (\%) & AUC & 8-bit AUC & 16-bit AUC & Compressed Model Size (MB) & 8-Bit  Model Size (MB) & 16-bit Model Size (MB) & Inference Time (ms) & 8-bit Inference Time (ms) & 16-Bit Inference Time (ms)\\
\hline
0.0 & 0.886 $\pm$ 0.000 & 0.887 $\pm$ 0.000 & 0.886 $\pm$ 0.000 & 22.739 $\pm$ 0.000 & 0.379 $\pm$ 0.000 & 9.382 $\pm$ 0.000 & 1.779 $\pm$ 0.130 & 1.388  $\pm$ 0.170 & 1.818 $\pm$ 0.124 \\
10.0 & 0.888 $\pm$ 0.003 & 0.888 $\pm$ 0.003 & 0.888 $\pm$ 0.003 & 21.575 $\pm$ 0.004 & 0.324 $\pm$ 0.011 & 9.062 $\pm$ 0.014 & 1.786 $\pm$ 0.145 & 1.427  $\pm$ 0.156 & 1.818 $\pm$ 0.150 \\
20.0 & 0.887 $\pm$ 0.004 & 0.887 $\pm$ 0.004 & 0.887 $\pm$ 0.004 & 19.903 $\pm$ 0.008 & 0.294 $\pm$ 0.018 & 8.446 $\pm$ 0.021 & 1.858 $\pm$ 0.164 & 1.373  $\pm$ 0.119 & 1.771 $\pm$ 0.156 \\
30.0 & 0.886 $\pm$ 0.004 & 0.886 $\pm$ 0.004 & 0.886 $\pm$ 0.004 & 18.041 $\pm$ 0.012 & 0.267 $\pm$ 0.017 & 7.749 $\pm$ 0.053 & 1.845 $\pm$ 0.167 & 1.359  $\pm$ 0.059 & 1.918 $\pm$ 0.217 \\
40.0 & 0.884 $\pm$ 0.005 & 0.884 $\pm$ 0.005 & 0.884 $\pm$ 0.005 & 16.053 $\pm$ 0.016 & 0.248 $\pm$ 0.016 & 6.966 $\pm$ 0.046 & 1.805 $\pm$ 0.138 & 1.389  $\pm$ 0.137 & 1.849 $\pm$ 0.201 \\
50.0 & 0.886 $\pm$ 0.004 & 0.886 $\pm$ 0.004 & 0.886 $\pm$ 0.004 & 14.002 $\pm$ 0.024 & 0.240 $\pm$ 0.015 & 6.350 $\pm$ 0.048 & 1.846 $\pm$ 0.154 & 1.363  $\pm$ 0.059 & 1.824 $\pm$ 0.184 \\
55.0 & 0.882 $\pm$ 0.005 & 0.882 $\pm$ 0.005 & 0.882 $\pm$ 0.005 & 12.878 $\pm$ 0.041 & 0.225 $\pm$ 0.014 & 5.897 $\pm$ 0.094 & 1.877 $\pm$ 0.165 & 1.353  $\pm$ 0.049 & 1.875 $\pm$ 0.214 \\
60.0 & 0.881 $\pm$ 0.005 & 0.881 $\pm$ 0.005 & 0.881 $\pm$ 0.005 & 11.762 $\pm$ 0.055 & 0.220 $\pm$ 0.014 & 5.510 $\pm$ 0.081 & 1.873 $\pm$ 0.180 & 1.420  $\pm$ 0.128 & 1.845 $\pm$ 0.143 \\
65.0 & 0.882 $\pm$ 0.004 & 0.882 $\pm$ 0.004 & 0.882 $\pm$ 0.004 & 10.604 $\pm$ 0.027 & 0.213 $\pm$ 0.014 & 5.066 $\pm$ 0.046 & 1.855 $\pm$ 0.169 & 1.361  $\pm$ 0.052 & 1.878 $\pm$ 0.192 \\
70.0 & 0.878 $\pm$ 0.005 & 0.878 $\pm$ 0.005 & 0.878 $\pm$ 0.005 & 9.395 $\pm$ 0.041 & 0.205 $\pm$ 0.013 & 4.583 $\pm$ 0.061 & 1.857 $\pm$ 0.198 & 1.397  $\pm$ 0.181 & 1.853 $\pm$ 0.173 \\
75.0 & 0.874 $\pm$ 0.005 & 0.875 $\pm$ 0.005 & 0.874 $\pm$ 0.005 & 8.147 $\pm$ 0.033 & 0.203 $\pm$ 0.014 & 4.084 $\pm$ 0.049 & 1.812 $\pm$ 0.134 & 1.378  $\pm$ 0.149 & 1.797 $\pm$ 0.131 \\
80.0 & 0.874 $\pm$ 0.005 & 0.874 $\pm$ 0.004 & 0.874 $\pm$ 0.005 & 6.796 $\pm$ 0.104 & 0.207 $\pm$ 0.038 & 3.554 $\pm$ 0.048 & 1.882 $\pm$ 0.165 & 1.397  $\pm$ 0.129 & 1.861 $\pm$ 0.210 \\
85.0 & 0.875 $\pm$ 0.005 & 0.875 $\pm$ 0.005 & 0.875 $\pm$ 0.005 & 5.427 $\pm$ 0.094 & 0.204 $\pm$ 0.019 & 3.021 $\pm$ 0.041 & 1.836 $\pm$ 0.153 & 1.398  $\pm$ 0.195 & 1.787 $\pm$ 0.135 \\
90.0 & 0.874 $\pm$ 0.004 & 0.874 $\pm$ 0.004 & 0.874 $\pm$ 0.004 & 4.005 $\pm$ 0.078 & 0.204 $\pm$ 0.014 & 2.383 $\pm$ 0.043 & 1.838 $\pm$ 0.161 & 1.379  $\pm$ 0.065 & 1.778 $\pm$ 0.165 \\
95.0 & 0.874 $\pm$ 0.004 & 0.874 $\pm$ 0.004 & 0.874 $\pm$ 0.004 & 2.596 $\pm$ 0.027 & 0.193 $\pm$ 0.012 & 1.708 $\pm$ 0.018 & 1.806 $\pm$ 0.162 & 1.371  $\pm$ 0.110 & 1.838 $\pm$ 0.146 \\
99.0 & 0.833 $\pm$ 0.015 & 0.833 $\pm$ 0.015 & 0.833 $\pm$ 0.015 & 1.748 $\pm$ 0.026 & 0.073 $\pm$ 0.010 & 1.281 $\pm$ 0.022 & 1.842 $\pm$ 0.209 & 1.428  $\pm$ 0.175 & 1.879 $\pm$ 0.177 \\
99.9 & 0.559 $\pm$ 0.118 & 0.559 $\pm$ 0.120 & 0.559 $\pm$ 0.118 & 1.441 $\pm$ 0.018 & 0.014 $\pm$ 0.004 & 1.096 $\pm$ 0.020 & 1.887 $\pm$ 0.203 & 1.436  $\pm$ 0.207 & 1.827 $\pm$ 0.188 \\
\end{tabular}}
\end{sc}
\end{small}

\caption{Results of pruning, with constant sparsity, and quantisation experiments for the CNN model.}
\label{tab:cnn_const_sparsity}
\end{center}

\begin{center}
\begin{small}
\begin{sc}
\scalebox{0.45}{%
\begin{tabular}{cccccccccc}
\hline
Sparsity (\%) & AUC & 8-bit AUC & 16-bit AUC & Compressed Model Size (MB) & 8-Bit  Model Size (MB) & 16-bit Model Size (MB) & Inference Time (ms) & 8-bit Inference Time (ms) & 16-Bit Inference Time (ms)\\
\hline
0.0 & 0.886 $\pm$ 0.000 & 0.887 $\pm$ 0.000 & 0.886 $\pm$ 0.000 & 22.739 $\pm$ 0.000 & 0.379 $\pm$ 0.000 & 9.382 $\pm$ 0.000 & 1.839 $\pm$ 0.115 & 1.365  $\pm$ 0.059 & 1.822 $\pm$ 0.115 \\
10.0 & 0.888 $\pm$ 0.003 & 0.888 $\pm$ 0.003 & 0.888 $\pm$ 0.003 & 21.460 $\pm$ 0.005 & 0.330 $\pm$ 0.010 & 9.272 $\pm$ 0.012 & 1.873 $\pm$ 0.104 & 1.376  $\pm$ 0.144 & 1.785 $\pm$ 0.117 \\
20.0 & 0.887 $\pm$ 0.004 & 0.887 $\pm$ 0.004 & 0.887 $\pm$ 0.004 & 19.818 $\pm$ 0.011 & 0.299 $\pm$ 0.010 & 8.783 $\pm$ 0.009 & 1.831 $\pm$ 0.181 & 1.350  $\pm$ 0.058 & 1.828 $\pm$ 0.147 \\
30.0 & 0.887 $\pm$ 0.004 & 0.887 $\pm$ 0.004 & 0.887 $\pm$ 0.004 & 18.018 $\pm$ 0.014 & 0.272 $\pm$ 0.010 & 8.163 $\pm$ 0.013 & 1.768 $\pm$ 0.143 & 1.375  $\pm$ 0.156 & 1.777 $\pm$ 0.132 \\
40.0 & 0.885 $\pm$ 0.004 & 0.885 $\pm$ 0.004 & 0.885 $\pm$ 0.004 & 16.079 $\pm$ 0.016 & 0.261 $\pm$ 0.008 & 7.462 $\pm$ 0.011 & 1.839 $\pm$ 0.144 & 1.425  $\pm$ 0.180 & 1.852 $\pm$ 0.122 \\
50.0 & 0.880 $\pm$ 0.005 & 0.880 $\pm$ 0.004 & 0.880 $\pm$ 0.005 & 14.018 $\pm$ 0.021 & 0.253 $\pm$ 0.010 & 6.687 $\pm$ 0.009 & 1.820 $\pm$ 0.096 & 1.371  $\pm$ 0.057 & 1.951 $\pm$ 0.189 \\
55.0 & 0.882 $\pm$ 0.006 & 0.882 $\pm$ 0.006 & 0.882 $\pm$ 0.005 & 12.928 $\pm$ 0.025 & 0.247 $\pm$ 0.009 & 6.269 $\pm$ 0.011 & 1.814 $\pm$ 0.132 & 1.349  $\pm$ 0.068 & 1.795 $\pm$ 0.144 \\
60.0 & 0.885 $\pm$ 0.004 & 0.885 $\pm$ 0.004 & 0.885 $\pm$ 0.004 & 11.827 $\pm$ 0.032 & 0.237 $\pm$ 0.010 & 5.837 $\pm$ 0.016 & 1.881 $\pm$ 0.155 & 1.388  $\pm$ 0.175 & 1.824 $\pm$ 0.144 \\
65.0 & 0.886 $\pm$ 0.005 & 0.886 $\pm$ 0.005 & 0.886 $\pm$ 0.005 & 10.629 $\pm$ 0.029 & 0.237 $\pm$ 0.008 & 5.356 $\pm$ 0.015 & 1.816 $\pm$ 0.111 & 1.393  $\pm$ 0.094 & 1.872 $\pm$ 0.114 \\
70.0 & 0.884 $\pm$ 0.004 & 0.884 $\pm$ 0.004 & 0.884 $\pm$ 0.004 & 9.454 $\pm$ 0.027 & 0.226 $\pm$ 0.010 & 4.866 $\pm$ 0.013 & 1.832 $\pm$ 0.131 & 1.413  $\pm$ 0.189 & 1.888 $\pm$ 0.209 \\
75.0 & 0.877 $\pm$ 0.006 & 0.877 $\pm$ 0.005 & 0.877 $\pm$ 0.006 & 8.188 $\pm$ 0.035 & 0.224 $\pm$ 0.012 & 4.312 $\pm$ 0.022 & 1.886 $\pm$ 0.159 & 1.377  $\pm$ 0.125 & 1.862 $\pm$ 0.118 \\
80.0 & 0.874 $\pm$ 0.004 & 0.874 $\pm$ 0.004 & 0.874 $\pm$ 0.004 & 6.869 $\pm$ 0.029 & 0.225 $\pm$ 0.010 & 3.724 $\pm$ 0.019 & 1.812 $\pm$ 0.131 & 1.354  $\pm$ 0.055 & 1.851 $\pm$ 0.104 \\
85.0 & 0.876 $\pm$ 0.006 & 0.876 $\pm$ 0.006 & 0.876 $\pm$ 0.006 & 5.544 $\pm$ 0.042 & 0.224 $\pm$ 0.010 & 3.114 $\pm$ 0.027 & 1.817 $\pm$ 0.085 & 1.389  $\pm$ 0.120 & 1.865 $\pm$ 0.184 \\
90.0 & 0.875 $\pm$ 0.006 & 0.875 $\pm$ 0.006 & 0.875 $\pm$ 0.006 & 4.184 $\pm$ 0.035 & 0.225 $\pm$ 0.011 & 2.489 $\pm$ 0.026 & 1.805 $\pm$ 0.156 & 1.374  $\pm$ 0.137 & 1.819 $\pm$ 0.168 \\
95.0 & 0.867 $\pm$ 0.005 & 0.868 $\pm$ 0.005 & 0.867 $\pm$ 0.005 & 2.757 $\pm$ 0.030 & 0.215 $\pm$ 0.008 & 1.826 $\pm$ 0.015 & 1.875 $\pm$ 0.133 & 1.340  $\pm$ 0.070 & 1.852 $\pm$ 0.125 \\
99.0 & 0.813 $\pm$ 0.029 & 0.813 $\pm$ 0.028 & 0.813 $\pm$ 0.029 & 1.834 $\pm$ 0.015 & 0.117 $\pm$ 0.002 & 1.379 $\pm$ 0.011 & 1.858 $\pm$ 0.152 & 1.372  $\pm$ 0.071 & 1.822 $\pm$ 0.138 \\
99.9 & 0.572 $\pm$ 0.093 & 0.574 $\pm$ 0.093 & 0.572 $\pm$ 0.093 & 1.631 $\pm$ 0.012 & 0.038 $\pm$ 0.001 & 1.260 $\pm$ 0.010 & 1.812 $\pm$ 0.146 & 1.338  $\pm$ 0.105 & 1.837 $\pm$ 0.170 \\
\end{tabular}}
\end{sc}
\end{small}
\caption{Results of pruning, with polyomial decay, and quantisation experiments for the CNN model.}
\label{tab:cnn_poly_decay}
\end{center}

\begin{center}
\begin{small}
\begin{sc}
\scalebox{0.45}{%
\begin{tabular}{cccccccccc}
\hline
Sparsity (\%) & AUC & 8-bit AUC & 16-bit AUC & Compressed Model Size (MB) & 8-Bit  Model Size (MB) & 16-bit Model Size (MB) & Inference Time (ms) & 8-bit Inference Time (ms) & 16-Bit Inference Time (ms)\\
\hline
0.0 & 0.725 $\pm$ 0.000 & 0.725 $\pm$ 0.000 & 0.725 $\pm$ 0.000 & 39.877 $\pm$ 0.000 & 8.514 $\pm$ 0.000 & 19.775 $\pm$ 0.000 & 104.745 $\pm$ 7.205 & 57.822  $\pm$ 2.065 & 101.097 $\pm$ 6.044 \\
10.0 & 0.713 $\pm$ 0.005 & 0.714 $\pm$ 0.005 & 0.713 $\pm$ 0.005 & 37.592 $\pm$ 0.001 & 8.476 $\pm$ 0.055 & 19.182 $\pm$ 0.001 & 101.537 $\pm$ 6.483 & 58.331  $\pm$ 4.018 & 102.999 $\pm$ 6.359 \\
20.0 & 0.713 $\pm$ 0.009 & 0.713 $\pm$ 0.009 & 0.712 $\pm$ 0.009 & 34.685 $\pm$ 0.004 & 8.304 $\pm$ 0.055 & 18.058 $\pm$ 0.003 & 100.002 $\pm$ 3.635 & 58.913  $\pm$ 2.192 & 100.789 $\pm$ 5.244 \\
30.0 & 0.716 $\pm$ 0.005 & 0.716 $\pm$ 0.005 & 0.716 $\pm$ 0.005 & 31.426 $\pm$ 0.017 & 7.866 $\pm$ 0.028 & 16.669 $\pm$ 0.013 & 99.027 $\pm$ 4.631 & 58.372  $\pm$ 2.317 & 102.448 $\pm$ 2.589 \\
40.0 & 0.717 $\pm$ 0.003 & 0.717 $\pm$ 0.003 & 0.717 $\pm$ 0.003 & 27.968 $\pm$ 0.018 & 7.249 $\pm$ 0.035 & 15.104 $\pm$ 0.013 & 99.339 $\pm$ 3.526 & 59.715  $\pm$ 2.995 & 98.867 $\pm$ 3.111 \\
50.0 & 0.717 $\pm$ 0.004 & 0.717 $\pm$ 0.003 & 0.717 $\pm$ 0.004 & 24.326 $\pm$ 0.060 & 6.488 $\pm$ 0.031 & 13.374 $\pm$ 0.037 & 100.159 $\pm$ 1.707 & 57.253  $\pm$ 3.243 & 98.758 $\pm$ 2.454 \\
55.0 & 0.717 $\pm$ 0.003 & 0.717 $\pm$ 0.003 & 0.717 $\pm$ 0.003 & 22.506 $\pm$ 0.028 & 6.043 $\pm$ 0.022 & 12.501 $\pm$ 0.016 & 100.154 $\pm$ 3.531 & 57.581  $\pm$ 2.160 & 99.826 $\pm$ 1.344 \\
60.0 & 0.716 $\pm$ 0.003 & 0.716 $\pm$ 0.003 & 0.716 $\pm$ 0.003 & 20.570 $\pm$ 0.057 & 5.600 $\pm$ 0.011 & 11.547 $\pm$ 0.038 & 96.827 $\pm$ 2.253 & 59.120  $\pm$ 3.938 & 100.417 $\pm$ 2.757 \\
65.0 & 0.716 $\pm$ 0.002 & 0.716 $\pm$ 0.002 & 0.716 $\pm$ 0.002 & 18.661 $\pm$ 0.046 & 5.109 $\pm$ 0.020 & 10.602 $\pm$ 0.032 & 99.359 $\pm$ 3.956 & 56.930  $\pm$ 1.875 & 98.610 $\pm$ 3.286 \\
70.0 & 0.721 $\pm$ 0.003 & 0.721 $\pm$ 0.003 & 0.721 $\pm$ 0.003 & 16.610 $\pm$ 0.040 & 4.611 $\pm$ 0.009 & 9.560 $\pm$ 0.026 & 99.850 $\pm$ 3.619 & 58.088  $\pm$ 2.586 & 99.487 $\pm$ 3.705 \\
75.0 & 0.720 $\pm$ 0.004 & 0.720 $\pm$ 0.004 & 0.720 $\pm$ 0.004 & 14.571 $\pm$ 0.080 & 4.086 $\pm$ 0.017 & 8.525 $\pm$ 0.054 & 99.100 $\pm$ 3.001 & 56.838  $\pm$ 1.251 & 101.674 $\pm$ 4.182 \\
80.0 & 0.720 $\pm$ 0.005 & 0.720 $\pm$ 0.005 & 0.720 $\pm$ 0.005 & 12.456 $\pm$ 0.037 & 3.492 $\pm$ 0.024 & 7.435 $\pm$ 0.025 & 100.223 $\pm$ 5.636 & 57.974  $\pm$ 3.011 & 101.821 $\pm$ 4.951 \\
85.0 & 0.718 $\pm$ 0.005 & 0.718 $\pm$ 0.005 & 0.718 $\pm$ 0.005 & 10.233 $\pm$ 0.032 & 2.868 $\pm$ 0.022 & 6.271 $\pm$ 0.024 & 99.288 $\pm$ 4.521 & 57.426  $\pm$ 2.333 & 101.591 $\pm$ 4.453 \\
90.0 & 0.712 $\pm$ 0.008 & 0.712 $\pm$ 0.008 & 0.712 $\pm$ 0.008 & 7.963 $\pm$ 0.015 & 2.147 $\pm$ 0.018 & 5.072 $\pm$ 0.011 & 99.541 $\pm$ 3.767 & 57.966  $\pm$ 2.890 & 100.404 $\pm$ 2.120 \\
95.0 & 0.709 $\pm$ 0.005 & 0.709 $\pm$ 0.005 & 0.709 $\pm$ 0.005 & 5.579 $\pm$ 0.010 & 1.281 $\pm$ 0.009 & 3.772 $\pm$ 0.008 & 100.564 $\pm$ 4.567 & 59.360  $\pm$ 4.570 & 105.974 $\pm$ 8.073 \\
99.0 & 0.627 $\pm$ 0.070 & 0.626 $\pm$ 0.070 & 0.627 $\pm$ 0.070 & 3.510 $\pm$ 0.022 & 0.349 $\pm$ 0.003 & 2.561 $\pm$ 0.014 & 106.898 $\pm$ 6.723 & 60.099  $\pm$ 3.057 & 105.807 $\pm$ 4.995 \\
99.9 & 0.500 $\pm$ 0.000 & 0.500 $\pm$ 0.000 & 0.500 $\pm$ 0.000 & 2.884 $\pm$ 0.021 & 0.073 $\pm$ 0.000 & 2.192 $\pm$ 0.016 & 106.033 $\pm$ 5.836 & 59.656  $\pm$ 1.634 & 103.384 $\pm$ 5.196 \\

\end{tabular}}
\end{sc}
\end{small}
\caption{Results of pruning, with constant decay, and quantisation experiments for the CNN-LSTM model.}
\label{tab:cnn_lstm_const_spar}
\end{center}

\begin{center}
\begin{small}
\begin{sc}
\scalebox{0.45}{%
\begin{tabular}{cccccccccc}
\hline
Sparsity (\%) & AUC & 8-bit AUC & 16-bit AUC & Compressed Model Size (MB) & 8-Bit  Model Size (MB) & 16-bit Model Size (MB) & Inference Time (ms) & 8-bit Inference Time (ms) & 16-Bit Inference Time (ms)\\
\hline
0.0 & 0.725 $\pm$ 0.000 & 0.725 $\pm$ 0.000 & 0.725 $\pm$ 0.000 & 39.877 $\pm$ 0.000 & 8.514 $\pm$ 0.000 & 19.775 $\pm$ 0.000 & 100.650 $\pm$ 6.398 & 59.763  $\pm$ 4.015 & 100.284 $\pm$ 5.564 \\
10.0 & 0.710 $\pm$ 0.008 & 0.710 $\pm$ 0.008 & 0.710 $\pm$ 0.008 & 37.570 $\pm$ 0.003 & 8.435 $\pm$ 0.075 & 19.157 $\pm$ 0.003 & 99.931 $\pm$ 4.931 & 62.185  $\pm$ 3.289 & 100.567 $\pm$ 5.565 \\
20.0 & 0.716 $\pm$ 0.005 & 0.716 $\pm$ 0.005 & 0.715 $\pm$ 0.005 & 34.575 $\pm$ 0.013 & 8.199 $\pm$ 0.078 & 17.958 $\pm$ 0.008 & 101.767 $\pm$ 7.881 & 59.693  $\pm$ 2.478 & 101.082 $\pm$ 5.767 \\
30.0 & 0.719 $\pm$ 0.002 & 0.719 $\pm$ 0.002 & 0.719 $\pm$ 0.002 & 31.296 $\pm$ 0.007 & 7.750 $\pm$ 0.058 & 16.534 $\pm$ 0.007 & 101.245 $\pm$ 7.045 & 58.906  $\pm$ 2.335 & 101.500 $\pm$ 6.486 \\
40.0 & 0.717 $\pm$ 0.003 & 0.717 $\pm$ 0.003 & 0.717 $\pm$ 0.003 & 27.830 $\pm$ 0.019 & 7.111 $\pm$ 0.032 & 14.976 $\pm$ 0.013 & 102.511 $\pm$ 6.836 & 60.104  $\pm$ 4.736 & 101.346 $\pm$ 6.560 \\
50.0 & 0.716 $\pm$ 0.004 & 0.716 $\pm$ 0.004 & 0.716 $\pm$ 0.004 & 24.260 $\pm$ 0.022 & 6.356 $\pm$ 0.023 & 13.272 $\pm$ 0.013 & 101.129 $\pm$ 6.275 & 56.916  $\pm$ 1.461 & 101.810 $\pm$ 4.067 \\
55.0 & 0.720 $\pm$ 0.003 & 0.720 $\pm$ 0.003 & 0.720 $\pm$ 0.003 & 22.469 $\pm$ 0.008 & 5.965 $\pm$ 0.022 & 12.417 $\pm$ 0.005 & 99.152 $\pm$ 4.247 & 58.770  $\pm$ 3.126 & 101.487 $\pm$ 5.960 \\
60.0 & 0.720 $\pm$ 0.002 & 0.720 $\pm$ 0.002 & 0.720 $\pm$ 0.003 & 20.509 $\pm$ 0.048 & 5.505 $\pm$ 0.023 & 11.444 $\pm$ 0.032 & 101.708 $\pm$ 6.309 & 57.557  $\pm$ 1.114 & 102.582 $\pm$ 5.183 \\
65.0 & 0.721 $\pm$ 0.003 & 0.721 $\pm$ 0.003 & 0.721 $\pm$ 0.003 & 18.621 $\pm$ 0.036 & 5.031 $\pm$ 0.027 & 10.515 $\pm$ 0.023 & 101.959 $\pm$ 5.998 & 59.341  $\pm$ 4.162 & 101.410 $\pm$ 5.462 \\
70.0 & 0.717 $\pm$ 0.006 & 0.717 $\pm$ 0.006 & 0.717 $\pm$ 0.006 & 16.614 $\pm$ 0.044 & 4.538 $\pm$ 0.013 & 9.511 $\pm$ 0.029 & 100.548 $\pm$ 5.630 & 57.209  $\pm$ 1.441 & 101.171 $\pm$ 5.868 \\
75.0 & 0.720 $\pm$ 0.004 & 0.720 $\pm$ 0.004 & 0.720 $\pm$ 0.004 & 14.560 $\pm$ 0.036 & 3.998 $\pm$ 0.022 & 8.468 $\pm$ 0.024 & 102.255 $\pm$ 7.137 & 60.227  $\pm$ 4.921 & 107.698 $\pm$ 11.603 \\
80.0 & 0.722 $\pm$ 0.002 & 0.722 $\pm$ 0.002 & 0.722 $\pm$ 0.002 & 12.403 $\pm$ 0.065 & 3.417 $\pm$ 0.015 & 7.351 $\pm$ 0.044 & 99.219 $\pm$ 5.613 & 60.596  $\pm$ 7.462 & 100.515 $\pm$ 6.432 \\
85.0 & 0.720 $\pm$ 0.006 & 0.720 $\pm$ 0.006 & 0.720 $\pm$ 0.006 & 10.202 $\pm$ 0.050 & 2.778 $\pm$ 0.013 & 6.211 $\pm$ 0.035 & 101.923 $\pm$ 6.068 & 59.312  $\pm$ 4.146 & 99.576 $\pm$ 6.652 \\
90.0 & 0.715 $\pm$ 0.003 & 0.715 $\pm$ 0.003 & 0.715 $\pm$ 0.003 & 7.938 $\pm$ 0.013 & 2.062 $\pm$ 0.011 & 5.035 $\pm$ 0.009 & 101.322 $\pm$ 8.083 & 58.816  $\pm$ 2.562 & 99.469 $\pm$ 6.873 \\
95.0 & 0.712 $\pm$ 0.012 & 0.712 $\pm$ 0.012 & 0.712 $\pm$ 0.012 & 5.551 $\pm$ 0.014 & 1.210 $\pm$ 0.002 & 3.729 $\pm$ 0.010 & 101.901 $\pm$ 6.458 & 59.531  $\pm$ 2.871 & 100.507 $\pm$ 6.143 \\
99.0 & 0.655 $\pm$ 0.079 & 0.653 $\pm$ 0.079 & 0.656 $\pm$ 0.079 & 3.535 $\pm$ 0.025 & 0.338 $\pm$ 0.001 & 2.552 $\pm$ 0.016 & 100.148 $\pm$ 5.250 & 59.986  $\pm$ 3.277 & 99.310 $\pm$ 6.912 \\
99.9 & 0.500 $\pm$ 0.000 & 0.500 $\pm$ 0.000 & 0.500 $\pm$ 0.000 & 2.932 $\pm$ 0.013 & 0.072 $\pm$ 0.000 & 2.196 $\pm$ 0.015 & 104.087 $\pm$ 7.670 & 51.875  $\pm$ 18.571 & 104.328 $\pm$ 8.254 \\
\end{tabular}}
\end{sc}
\end{small}
\caption{Results of pruning, with polyomial decay, and quantisation experiments for the CNN-LSTM model.}
\label{tab:cnn_lstm_poly_decay}
\end{center}
\end{table*}

\end{document}